# CFD modeling of hydrogen release and dispersion in a congested container

Hector Amino [1, 2], Lynda Porcheron [1], Jerome Daubech [3], Emilie Ricrot [4], Annabelle Brisse [4], Emmanuel Leprette [3], Olivier Hurisse [1]

[1] Fluid Mechanics, Energy and Environment Department, EDF R&D, Chatou, France
[2] CEREA, Ecole des Ponts, EDF R&D, Champs-sur-Marne, France
[3] INERIS, BP 2, 60550, Verneuil-en-Halatte, France
[4] HYNAMICS, 8-10 Avenue de l'Arche Immeuble Colisée Gardens, 92400 Courbevoie, France

ABSTRACT

Hydrogen plays an important role in driving decarbonization within the current global energy landscape. As hydrogen infrastructures rapidly expand beyond their traditional applications, there is a need for comprehensive safety practices, solutions, and regulations. Within this framework, the dispersion of hydrogen in enclosed facilities presents a significant safety concern due to its potential for explosive accidents. In this study, hydrogen dispersion in a confined and congested environment (37 m$^3$ container) is studied using computational fluid dynamics simulations. The experimental setup mirrors previous INERIS investigations, featuring a centrally located hydrogen injection point on the floor with a 20 mm diameter injector and a release rate of 35 g/s, resulting in a Froude number of 650. This yields an inertial jet and a challenging dispersion scenario for numerical prediction. Concentration mapping is carried out by 3 oxygen analyzers distributed throughout the 37 m3 chamber. Comparisons are made between the measured and numerical data to validate the solver used under such conditions. Best practice guidelines are followed, and sensitivity studies involving grid refinement and boundary conditions are conducted to ensure robust simulation results. The findings highlight the model's ability to reproduce the hydrogen concentration distribution for both empty and congested containers and underline the role of accounting for leakages in such scenarios.

## 1. INTRODUCTION

Hydrogen has become a promising alternative to fossil fuels, but its use comes with significant safety challenges. Its wide flammability range (4% to 75% in air), low ignition energy, high flame velocity, and low density make it prone to form flammable mixtures and disperse rapidly, leading to serious explosion hazards [1]. Hydrogen can cause various types of explosions, including deflagration, detonation, vapor cloud explosions and flash fires. Despite these risks, hydrogen's potential for clean energy makes it valuable in the global energy transition, necessitating advanced risk detection technologies and robust safety protocols to mitigate these dangers.

Over past, researchers and industry professionals assessed hydrogen related risks and mitigation strategies associated with its explosion. Key areas of focus include understanding the release of compressed gas, its dispersion in open and closed environments, the characteristics of deflagration, and the potential transition from deflagration to detonation (DDT) [1]. More specifically, a thorough understanding of hydrogen dispersion is crucial for designing safer facilities and implementing effective measures to prevent the accumulation of gas that could lead to explosive atmospheres.

In this context, a significant challenge is to characterise the formation of flammable clouds within potentially congested industrial geometries. Accidental events can result in gas releases with different dynamics; for example, a small release diameter produces a buoyant jet, whereas a clean break leads to a momentum jet with higher leak rates. The combination of complex geometrical configurations and significant release rates makes experimental feasibility on test benches difficult and may raise safety concerns.

Light gas dispersion in enclosed systems has been studied extensively (refer to Table 1 for a list of some experiments). However, typical gas leak rates for configurations above 10 m³ are usually below 10 g/s, resulting in lower jet Froude numbers and in slower dispersion. This type of release yields to stratified atmospheres, especially if there are obstacles present within the system. In addition, most studies include vents in the system, reflecting the industrial application to vented configurations.

Table 1. Example of experimental enclosed light gas dispersion done in the past.

| Ref. | Year | Volume [m³] | Inj. Mass flow [g/s] | Froude number | Congestion |
| --- | --- | --- | --- | --- | --- |
| Pitts et al. [2] | 2009 | ~1.7 | ~[9, 23] | ~[30, 73] | / |
| Gupta et al. [3] | 2009 | ~41.0 | ~[0.002, 2.5] | ~[0.8, 31] | / |
| Merilo et al. [4] | 2011 | ~60.2 | ~[0.29, 3.1] | ~[62, 648] | Vehicle |
| Lacome et al. [5] | 2011 | 80 | ~[0.2, 1] | ~[6.8, 211] | / |
| Pitts et al. [6] | 2012 |  | ~1.35 | ~38 | Vehicle |
| Cariteau and Tkatschenko [7] | 2013 | ~1.1 | ~[0.075, 0.68] | ~[33, 300] | / |

The mentioned experiments have been used for numerical validation, particularly for CFD software. For example, Giannissi et al. [8] investigated the performance of various turbulent approaches in CFX by reproducing data from [3]. Midha presented validation results in the dispersion framework [9] using the FLACS software, where high Reynolds jets were validated for unconfined configurations. Studies on impinging jets [10], the impact of wind on natural ventilation, and other specific physics [11, 12, 13] can also be highlighted (refer to [14] for an extensive list of hydrogen release CFD simulations). The validation of these methods allows for their use in parametric studies and the extraction of unmeasured data. For instance, integral turbulent quantities from flammable clouds can be retrieved from 3D results and applied in phenomenological models for explosion scenarios.

In this study, the dynamics of an inertial hydrogen release within a 37.5 m³ container are invest, both in its empty state and congested states. A vertical leak, characterized by a Froude number of 650, is simulated, similar to a clean break scenario, using the open-source software code_saturne [15]. The evolution of gas concentration is then compared with experimental measurements. Additionally, we conduct a mesh sensitivity analysis and assess the influence of potential container leaks under pressurized conditions on the results. The following sections provide a detailed account of the experimental and numerical setups and present a comparison between the simulation outcomes and the measured data.

2. EXPERIMENTAL SET-UP

The studied system is a 37.5 m³ container ([6 x 2.5 x 2.5] m, also studied for multi-vent deflagrations in [16, 17]), designed to withstand an explosion overpressure of 3 bar (see Fig. 1). Its structure is composed of H-iron and modular side frames that can accommodate vent panels, viewing windows or solid walls to block the given surface. Hydrogen (100%) is injected from a 5 m³ tank through a 20 mm diameter circular orifice positioned at the centre of the container floor. The injection pressure is monitored by a pressure sensor located upstream of the pilot valve. The gas mass flow rate is 35 g/s for 15 seconds for the empty container and only 5.5 seconds for the congested container. This flow rate is chosen to homogenize the flammable atmosphere without additional mechanical momentum sources, enabling an experimental deflagration to occur shortly after. Gas concentration is measured at three locations using oxygen analysers, with sensors coordinates listed in Table 2. Note that in the configuration where the chamber was empty, a significant delay—on the order of 30 seconds—was observed between the response times of the upper and lower sensors compared to the middle sensoris discrepancy was not present when the chamber was obstructed. Given that the gas concentration sensors operate via a sampling pump, a difference in the length of the sampling lines could have explained the observed delay. However, all sampling lines were of identical length, ruling out this factor. Furthermore, the same middle sensor was used in both configurations (empty and obstructed chamber), eliminating the possibility of sensor malfunction or variability. The middle sensor is in proximity to the injection point. It was hypothesized that a localized pressure drop at the sampling point could have temporarily affected the measurement. Nevertheless, since the release lasted only 15 seconds, any associated pressure effect would have ceased once the leak ended, and the sensor would have registered a response earlier, not later. A plausible explanation for the delayed response is the presence of a localized restriction—such as a partial occlusion or pinch—in the sampling line, which would have reduced the internal cross-sectional area and thereby delayed the transport of the gas sample. During the placement of the obstacles in the chamber, the sampling lines were repositioned. It is therefore likely that this restriction was inadvertently removed during reinstallation, explaining the disappearance of the delay in the obstructed configuration.

During the tests, the system was sealed, with modular side frames fitted with solid blocks instead of plastic sheets at the vent locations. However, under high pressurization conditions, gaps of 2 mm wide may appear along the edges of the viewing windows (a total of 9 frames, including 3 in the ceiling).

In the congested scenario, two pipe racks are installed inside the container. Pipes with a diameter of 10 cm and a length of 1.5 m are secured in two supports, each measuring [1.05 x 0.05 x 1.5] m, as shown in Fig. 2 (right). Both structures are mounted on two beams, each measuring [3.6 x 0.18 x 0.4] m. Fig. 3 provides a top view of the congested system.

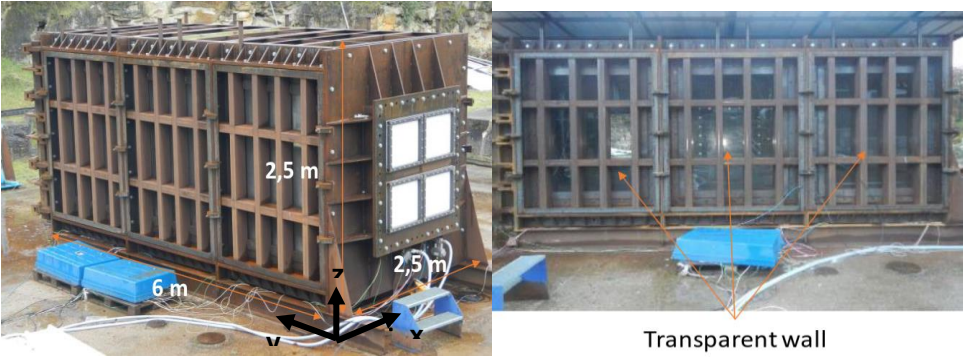

Figure 1. Illustration of the explosion chamber.

Table 2. Gas measurement sensor's location.

| Name | X (m) | Y (m) | Z (m) |
| --- | --- | --- | --- |
| Low | 0.3 | 5.7 | 0.2 |
| Medium | 1.25 | 2.5 | 1.0 |
| High | 2.1 | 0.4 | 2.25 |

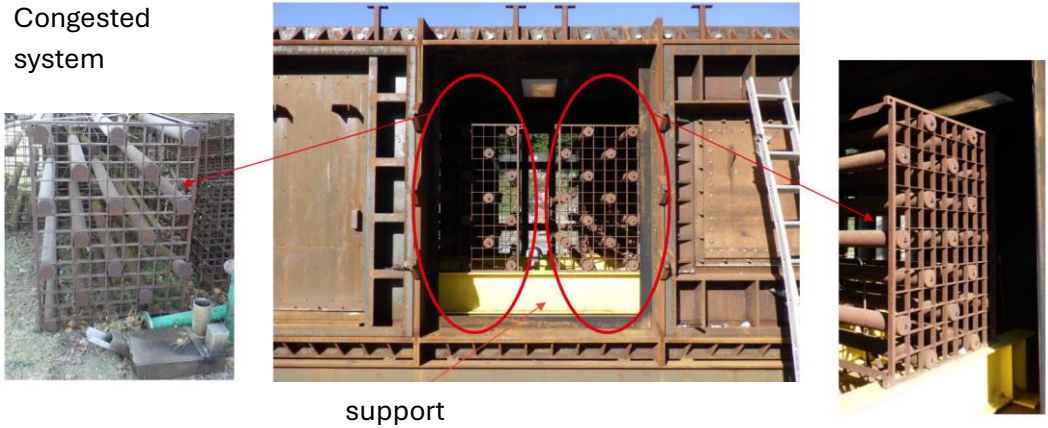

Figure 2. Congested container with the two pipe racks fixed to a support.

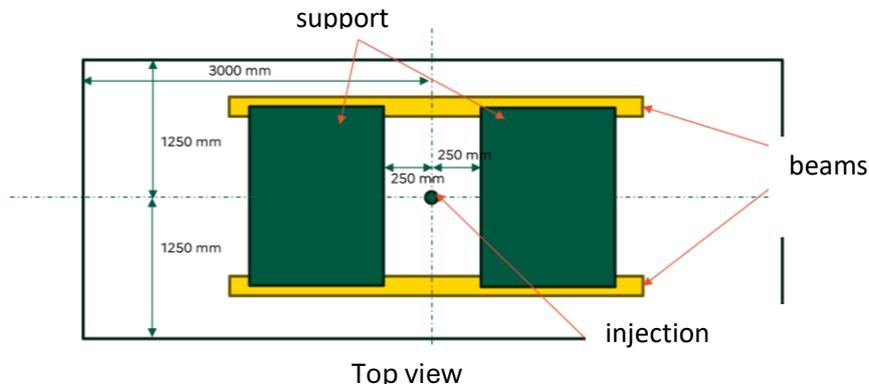

Figure 3. Top view of the congested system.

### 3. NUMERICAL SETUP

Based on the experimental work, a transient hydrogen release validation study is conducted using the CFD open-source solver code_saturne [15]. The configuration is discretised using a hexahedral mesh. Initially, the container is assumed to be fully closed, considering only the fluid volume inside. Then, the simulation domain includes the external environment, with a total open surface area of 0.15 m² (see Fig. 4, right). Given the lack of gas-tightness in the container, the occurrence of leakage at the junctions were accounted for by including slots at the junctions between the panels framing the side walls and ceiling. These slots were specifically placed in the lateral container walls and the ceiling zone (the blue circle in Fig. 4). In the second scenario, the fluid domain is larger than the container, covering a volume of 6 x 6.5 x 4 m.

Uniformly sized hexahedral cells, starting from a length of 5 cm, are used, resulting in reference meshes consisting of approximately 300,000 cells for the closed container and 1.2 million cells for the scenario with openings (see Fig. 4 and 5). Following best practices guidelines for this type of scenario, simulations are also performed using a refined mesh, where the cell dimensions for each axis are halved. The presence of obstacles is considered by defining solid volumes within the container, as illustrated in Fig. 5. Given the reference cell size, only solids with length above 5 cm are considered.

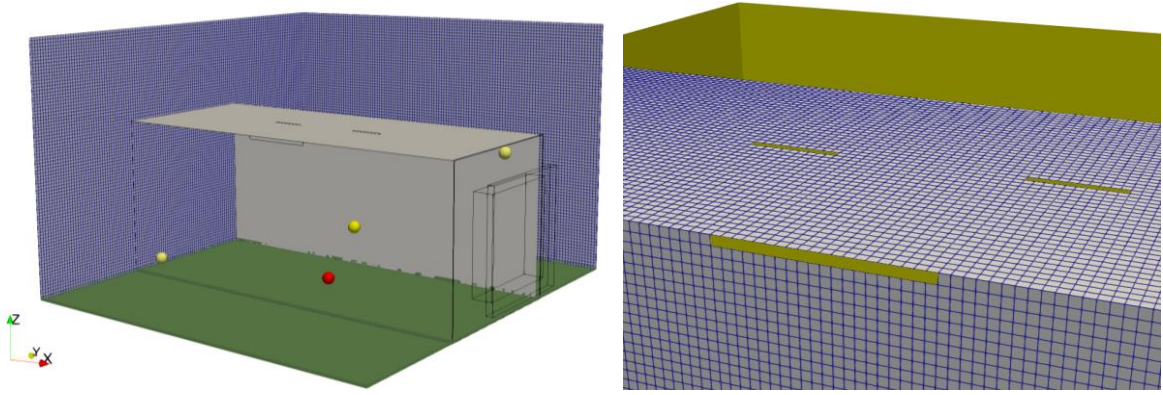

Figure 4. (Left) Numerical representation of the studied container without obstacles (dx = 5 cm). Yellow dots represent the monitoring probes positions; the red dot is the injection position. (Right) Zoom on the container openings. Backwalls are coloured in yellow for visualisation purposes.

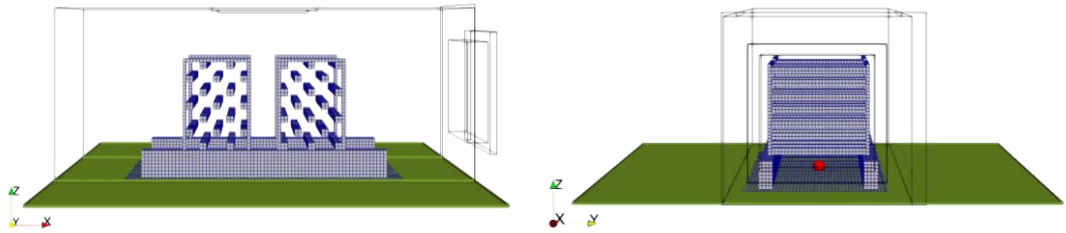

Figure 5. Congested system side and front views of the obstacles for the numerical simulation.

A prediction-correction compressible scheme [18] is used, in combination with a second-order spatial and time scheme. The RANS turbulence approach is used with the k-epsilon linear production model [19]. Such approach was already used in the literature for hydrogen or helium dispersion (e.g. [20], see [21] for a review of different studies using this closure). A mixture of ideal gases is used for the equation of state. Given the injector diameter of 20 mm, the notional nozzle approach is applied in this study [22]. Following this method, the gas release is numerically modelled as a volume source term in the injection zone. As previously mentioned, two scenarios are simulated for each container configuration. In the first scenario, where the container is closed, a non-slip wall condition is applied to all boundary faces, using a two-scale log law wall function for velocity and scalars. In the second scenario, outlets at the atmospheric hydrostatic pressure are defined at the domain borders, except for the container walls and the ground. Finally, the simulation time step is variable over time to preserve the Courant number under the unity. The transient dispersion flow is tracked using monitoring probes, and the local gas concentration values are compared with the experimental results.

### 4. RESULTS

In this section, the measured local concentrations are compared to the numerical results. For the empty and congested container, a mesh sensitivity analysis is performed as well as a comparison between a closed and an opened container. It should be noted that the curves of the experimental results have been shifted to the left to be consistent with the sampling time of the different probes during the tests (length of the pipes and residence time of the gases to the analyser). All measured data have the same initial acquisition time.

**Results for the empty container**

For the three probes, Fig. 6 shows the evolution of the hydrogen concentration during the leak for each closed system meshes. The numerical reproduction of the container gas filling aligns with the expected physics: a transient increase in concentration is observed during the leak, followed by a homogeneous value. However, the results globally overestimate the plateau concentration, which is expected to be around 12%. A threshold value around 15% is expected as a 15 s leak mass flow of 35 g/s yields to around 6.3 Nm$^3$. Such result can be explained by the presence of mixture leakage under the overpressure induced by the gas release. Additionally, the rise in hydrogen concentration at the medium sensor appears almost immediate in the numerical simulation, whereas it is observed after 30 seconds in the experiment. This discrepancy can be attributed to the time required for gas presence measurement in the system.

As the results are very similar for the two mesh sizes used, a mesh size of 5 cm will be used for the following calculations.

Due to the overprediction of concentration and the hypothesis of possible gas leak, a second simulation with a reduced release rate of 0.25 g/s was conducted. This rate was selected to achieve a concentration close to 12% over the same release time in a closed container. As shown in Fig. 7, this release rate yields to a maximum hydrogen concentration value that approaches the expected level. The numerical results also illustrate the propagation of the gas cloud within the container: an initial peak is observed at the low and high sensors, corresponding to the arrival of the gas front. The threshold value corresponds to the end of the release, and the consistent values across all sensors indicate a homogeneous mixture after 20 seconds.

The presented findings suggest that the measured concentration for the experimental release rate is only achieved if part of the mixture escapes through leaks in the container structure.

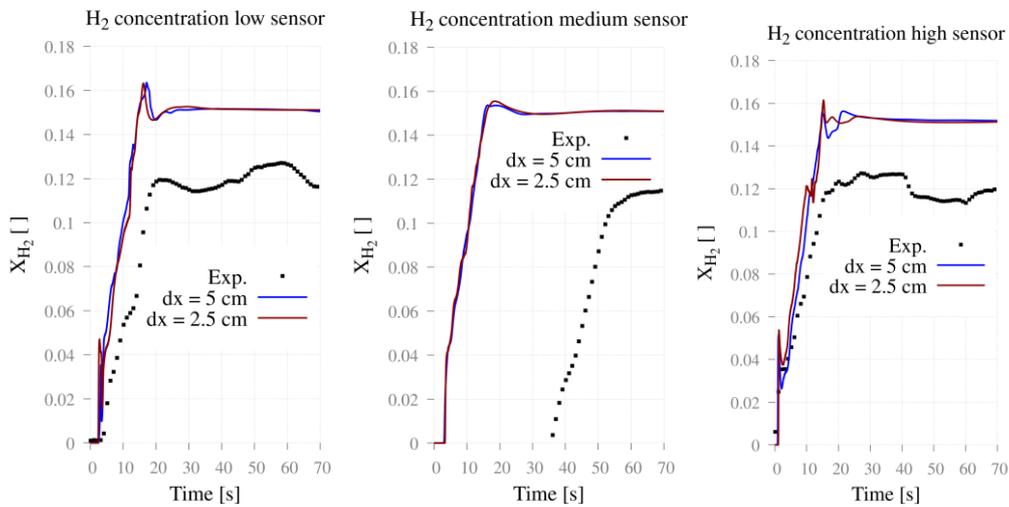

Figure 6. Closed system. Hydrogen concentration vs time for the empty container using an injection mass flow of 35 g/s. Mesh sensitivity analysis (2.5 and 5 cm cells) for a 15 s duration leak.

To assess the impact of potential leaks within the container, we estimated the surface area based on the studied configuration and performed the simulation of an extended system (illustrated in Fig. 4). As anticipated, the maximum concentration value decreased when small leaks in the upper part of the container were considered (Fig. 8). Consequently, the numerical results match better the measured concentrations, both for the plateau value and the slope of the gas increase. The release dynamics is close to an impinging jet, given an important jet velocity. After hitting the ceiling, the jet expands over the container walls, before reaching the ground and mixing. The dispersion dynamics creates recirculating zones, notably for height close to the ground and to the ceiling. This is illustrated later in Fig. 11.

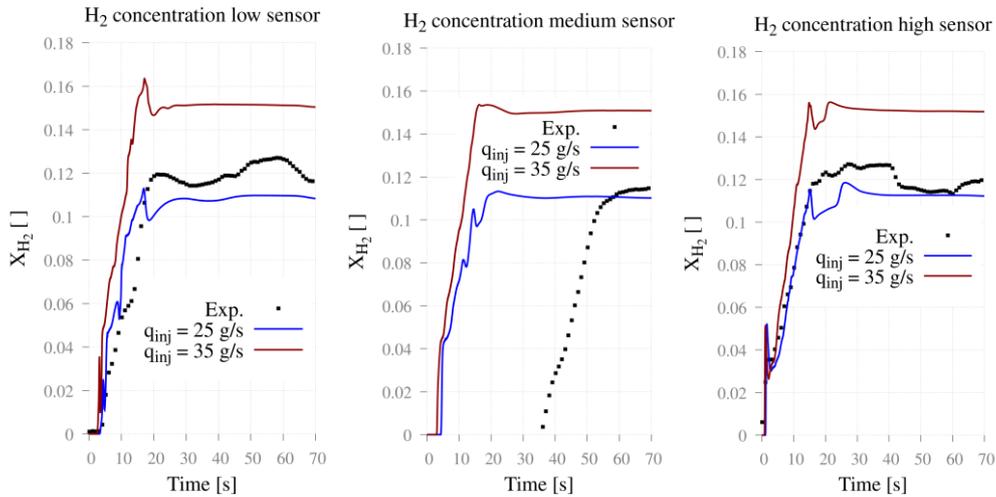

Figure 7. Closed system. Hydrogen concentration vs time for two injection mass flows (25 g/s and 35 g/s for a 15s duration leak).

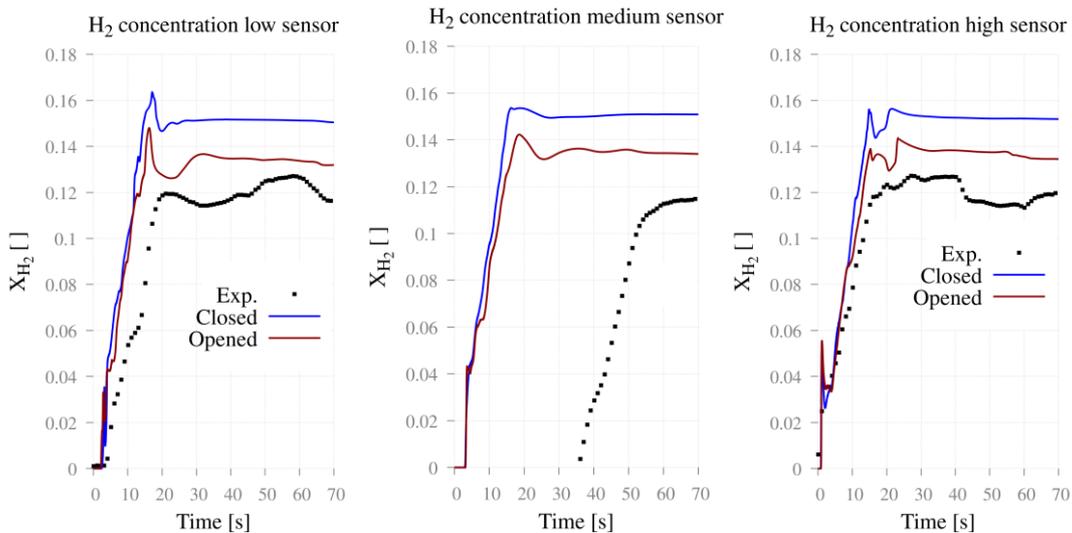

Figure 8. Hydrogen concentration vs time for a closed and opened container (35 g/s for a 15s duration leak).

Some error on the maximum concentration numerically predicted can be observed (around 14% instead of 12%). This can be explained by an underestimation of the container leakage surface. Also, the choice of the turbulent approach may impact the numerical results.

**Congested container**

This section presents the comparison between the numerical and experimental results for the congested system. As in the previous case, a mesh sensitivity analysis is performed for the closed container scenario (Fig. 9). Results show again a little results variability, leading to the use of the 5 cm cells mesh for the next simulations.

Overall, the numerical concentration profiles align well with the measured data. However, the RANS simulation fails to capture the hydrogen fluctuations observed experimentally. The LES approach should offer an accurate representation of jet break-up and interaction with obstacles. In this instance, the open system resulted in underestimated concentrations compared to the closed system (Fig. 10). This discrepancy could be attributed to the shorter leak duration, which is 9.5 seconds less than the previous scenario: the higher overpressure from a longer injection process likely increases leakiness.

Fig. 11 and 12 illustrate respectively cut planes of the hydrogen concentration and velocity fields at t = 5s and 20s. During the release, the high-speed jet reaches the ceiling and spreads through the walls. Some of the mixture is expelled through the openings due to dispersion dynamics, explaining the lower concentration measured at the monitoring locations. Following the release, a transition to a buoyant dynamic regime is observed, as shown in Fig. 12, along with the formation of a large recirculation zone around the obstacles and local flow recirculation within the obstacle zone.

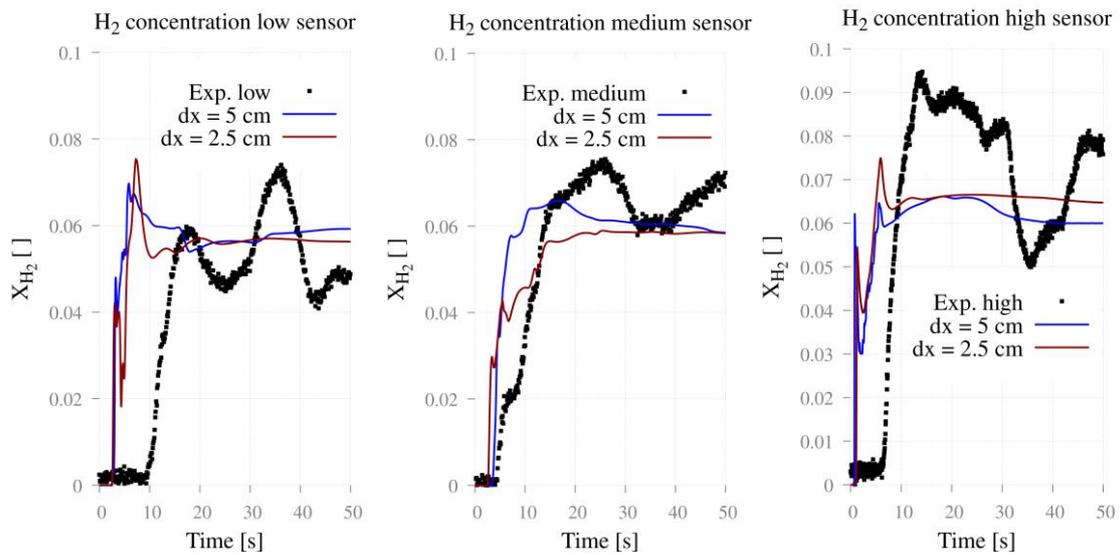

Figure 9. Closed system. Hydrogen concentration over time for the congested container using an injection mass flow of 35 g/s. Mesh sensitivity analysis.

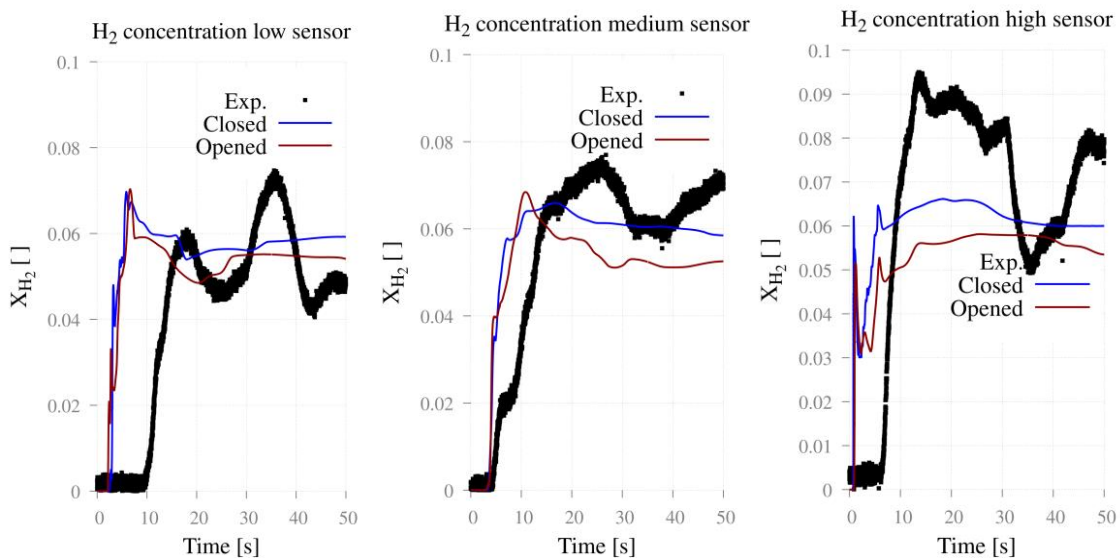

Figure 10. Simulation for the congested configuration. Hydrogen concentration vs time for a closed and opened container (35 g/s for a 5.5s duration leak).

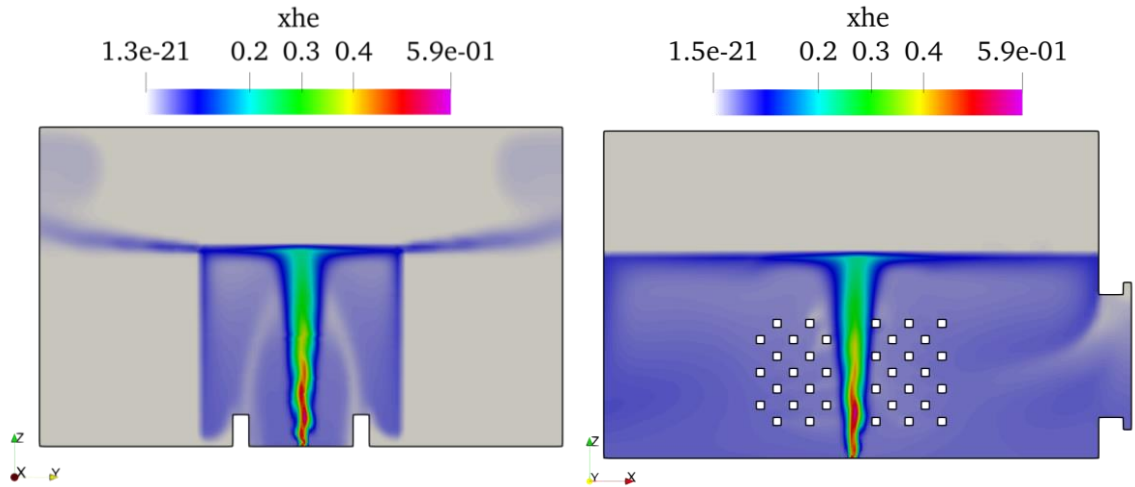

Figure 11. Simulation of the congested dispersion: 2D cut planes of the hydrogen concentration field at t = 5s.

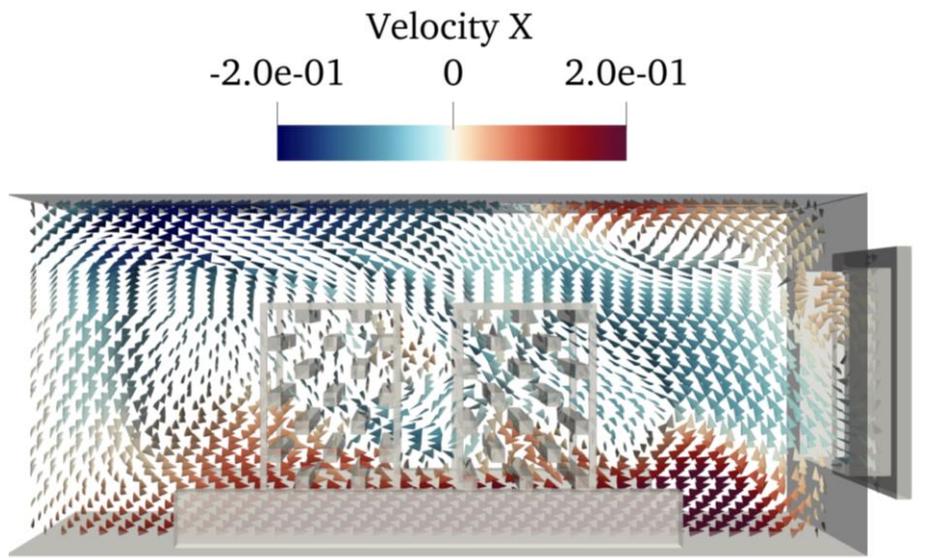

Figure 12. Numerical horizontal velocity field represented on arrow glyphs at t = 20s.

## 5. CONCLUSION

This study investigates the dynamics of inertial hydrogen release within a 37.5 m³ container, both in empty and congested states, using the open-source software code_saturne. The simulations, characterized by a Froude number of 650, were compared with experimental measurements to validate the numerical approach. The system was initially sealed, but potential leaks were considered under high pressurization conditions. The numerical results generally aligned with experimental data, though some discrepancies were noted, particularly in the immediate rise of hydrogen concentration and the overestimation of plateau values. The consideration of container leaks improved the accuracy of the simulations, notably for a longer release time.

Results demonstrate the capability of 3D simulations to predict high Froude number gas dispersion within a congested container, assuming a thorough understanding of the studied configuration. Future work should focus on testing the LES approach to predict the local concentration fluctuations observed in the measurements. Additionally, validation will be extended to include dynamic measurements, such as the velocity at various container locations. Results will be used to extract turbulent quantities for phenomenological models to predict the overpressure generated by an explosive scenario.